\documentclass[sigconf]{acmart}

\usepackage{subfigure}
\usepackage{graphicx}
\usepackage{caption}
\usepackage{tabularx}
\usepackage{multirow}

\newcolumntype{Z}{>{\raggedleft\let\newline\\\arraybackslash\hspace{0pt}}X}

\def\BibTeX{{\rm B\kern-.05em{\sc i\kern-.025em b}\kern-.08emT\kern-.1667em\lower.7ex\hbox{E}\kern-.125emX}}
    
\copyrightyear{2019} 
\acmYear{2019} 
\acmConference[IHC '19]{Proceedings of the 18th Brazilian Symposium on Human Factors in Computing Systems}{October 22--25, 2019}{Vit\'oria - ES, Brazil}
\acmBooktitle{Proceedings of the 18th Brazilian Symposium on Human Factors in Computing Systems (IHC '19), October 22--25, 2019, Vit\'oria - ES, Brazil}
\acmPrice{15.00}
\acmDOI{10.1145/3357155.3358441}
\acmISBN{978-1-4503-6971-8/19/10}

\begin{document}

%
\title[Characterising Volunteers' Task Execution Patterns Across Projects]{Characterising Volunteers' Task Execution Patterns Across Projects on Multi-Project Citizen Science Platforms}

%
\author{Lesandro Ponciano}
\orcid{0000-0002-5724-0094}
\affiliation{%
  \institution{Pontifical Catholic University of Minas Gerais}
  \streetaddress{Street Adress}
  \city{Belo Horizonte}
  \state{Minas Gerais, Brazil}
  \postcode{30535-901}}
\email{lesandrop@pucminas.br}

\author{Thiago Emmanuel Pereira}
\affiliation{%
  \institution{Federal University of Campina Grande}
  \streetaddress{Street Adress}
  \city{Campina Grande}
  \state{Paraiba, Brazil}
  \postcode{Post Code}}
\email{temmanuel@computacao.ufcg.edu.br}

%
\renewcommand{\shortauthors}{Lesandro Ponciano and Thiago Emmanuel Pereira}

%

\begin{abstract}

Citizen science projects engage people in activities that are part of a scientific research effort. On multi-project citizen science platforms, scientists can create projects consisting of tasks. Volunteers, in turn, participate in executing the project's tasks. Such type of platforms seeks to connect volunteers and scientists' projects, adding value to both. However, little is known about volunteer's cross-project engagement patterns and the benefits of such patterns for scientists and volunteers. This work proposes a Goal, Question, and Metric (GQM) approach to analyse volunteers' cross-project task execution patterns and employs the Semiotic Inspection Method (SIM) to analyse the communicability of the platform's cross-project features. In doing so, it investigates what are the features of platforms to foster volunteers' cross-project engagement, to what extent multi-project platforms facilitate the attraction of volunteers to perform tasks in new projects, and to what extent multi-project participation increases engagement on the platforms. Results from analyses on real platforms show that volunteers tend to explore multiple projects, but they perform tasks regularly in just a few of them; few projects attract much attention from volunteers; volunteers recruited from other projects on the platform tend to get more engaged than those recruited outside the platform. System inspection shows that platforms still lack personalised and explainable recommendations of projects and tasks. 
\end{abstract}

%
%


%

%
\maketitle

\section{Introduction}
Citizen science projects engage volunteers in scientific research effort~\cite{irwin1995citizen,wiggins2011conservation,preece2016citizen,eitzel2017citizen}. In crowd-sourced citizen science projects, volunteers participate via an online platform that ease the design, management and execution of the tasks that make up the projects~\cite{simpson2014zooniverse,ponciano2014considering}. Multi-project platforms, such as Zooniverse (\url{zooniverse.org}) and Crowdcrafting (\url{crowdcrafting.org}), allow scientists to host their projects and ask for contributions. A crowd of volunteers can join the platform, search for the available projects and contribute by performing tasks in the projects they are interested in. There are several types of projects and tasks that volunteers can perform. Some projects require only simple abilities, such as data collecting and reporting. Other projects require more complex cognitive abilities such as data aggregation and classification ~\cite{wiggins2011conservation,haklay2013citizen}, usually defined as human computation tasks~\cite{von2008human,law2011human,lintott2013human,ponciano2014considering,michelucci2016power}. The scientists involved in multi-project citizen science platforms seek both a continuous recruiting of new volunteers for their projects and to keep engaged those volunteers who already participate.

Multi-project platforms can foster the participation of the volunteers in projects, connecting scientists and volunteers, and providing benefit to both of them. Notwithstanding this potential, there are few studies on how the features provided by platforms to scientists and volunteers impact the effectiveness of the platforms. In particular, little is known about volunteers' cross-project engagement patterns and how the projects benefit from such patterns. This study aims \textit{at filling this gap by characterising how volunteers and scientists behave in multi-project platforms and the benefits they take from them}.  

Three research questions guide the study:
\begin{enumerate}
    \item to what extent the multi-project nature of platforms facilitate the attraction of volunteers to new projects; 
    \item to what extent the existence of multiple projects can lead volunteers to engage with the platform regularly;
    \item what are the main features of multi-project platforms fostering a cross-project engagement.
\end{enumerate}

In answering these questions, this study contributes to understanding the dynamics of participants' engagement, their experience, and identifies opportunities for designing more effective multi-project platforms.  

The method proposed and employed in this study\textit{ is designed to be used by researchers and platform managers to (1) elicit volunteers' cross-project recruitment and engagement, and (2) improve volunteers' cross-project experience}. The method combines a quantitative characterisation of volunteers' cross-project behaviour and a communicability inspection of platforms' cross-project features. The characterisation is based on the Goal, Question and Metric (GQM) approach. This approach is useful in guiding the analysis of data collected from systems by taking into account the \textit{goals} of the agents involved in the system, \textit{questions} associated with those goals, and \textit{metrics} that can be established to answer the questions and verify if the goal has been met~\cite{caldiera1994goal,van1999goal,van2014agile,suzuki2018interaction}. The communicability inspection is based on the Semiotic Inspection Method, which is relevant for the study of the signs used by the designers to communicate to the users the features of the system~\cite{de2005semiotic,de2006semiotic}. To the best of our knowledge, no previous study has employed GQM and Semiotic Inspection in the context of multi-project citizen science platforms.

To evaluate the applicability of the proposed method and gain some knowledge of the cross-project recruitment and engagement patterns on real platforms, we study five platforms: Crowdcrafting, Socientize, GeoTag-X, Zooniverse, and CitSci.org. 
The quantitative characterisation of volunteers' cross-project behaviour is performed with the typical data stored by the platforms. This data is made available publicly through their application programming interface (API). The communicability inspection is performed in the platforms' Web interface used by the volunteers. The proposed method and findings characterise individual multi-project platforms. They are not designed to compare different platforms.

The obtained results provide an overview of the design of multi-project platforms and the behaviour of volunteers on such platforms. The characterisation study shows that, while volunteers tend to explore various projects available on multi-project platforms, most of the volunteers regularly engage in just a few of them. Few projects on the platform attract much attention from volunteers. Volunteers recruited from other projects tend to get more engaged than those recruited outside the platform. Results from the communicability study show three classes of signs in platforms' interfaces that are associated to features to foster a cross-project engagement: 1) signs of project search; 2) signs of projects preferred or recommended by the platform; and 3) signs of participant's projects. These results show that platforms still lack personalised and explainable strategies for recommending projects that fit volunteers' preferences and behaviours.

\section{Background and Related Work}

The study of volunteers' cross-project engagement interrelates the subjects of volunteering, engagement, human computation and citizen science. Studies in several disciplines have addressed these subjects, such as psychology~\cite{clary1998understanding}, sociology~\cite{wilson2000volunteering}, and computer science~\cite{o2008user,law2011human}. The present work builds on a broad set of concepts, methods and knowledge accumulated by such studies. In this section, we present such background and discuss the related work. 

\subsection{Volunteering and Engagement}

The contributing behaviour of people taking part in citizen science platforms based on human computation can be examined in the light of volunteering and human engagement literatures~\cite{ponciano2014finding,west2016recruiting}. Volunteering literature usually distinguishes between two different types of participation behaviour: helping and volunteering~\cite{clary1998understanding,wilson2000volunteering}. Helping behaviour designates \textit{sporadic} participation in which the individual is faced with an unexpected request to help someone to do something. Volunteering behaviour designates \textit{planned} participation when people are actively seeking out opportunities to help others and commit themselves at a considerable personal cost. The literature on human engagement focuses on human behaviour when individuals self-invest personal resources such as cognitive power~\cite{o2008user}. O'Brien and Toms proposed an operational definition of engagement comprised of a point of engagement, and periods of sustained engagement, disengagement, and re-engagement~\cite{o2008user}. This definition of engagement has been considered in other studies~\cite{ponciano2014finding, boakes2016patterns} and is the basis of the engagement view employed in this study.

\subsection{Human Computation Tasks}

Volunteers in citizen science can perform a broad range of tasks. Some tasks require simple abilities, such as data collecting and reporting. On the other hand, some tasks require more complex cognitive abilities such as data aggregation and classification~\cite{wiggins2011conservation,wiggins2012goals,haklay2013citizen,ponciano2018agreement}. Therefore, various types of engagement are possible~\cite{wiggins2011conservation,preece2016citizen}. In crowd-sourced citizen science projects, volunteers engage cognitively by performing \textit{human computation tasks}, which are tasks that humans perform better than current computers. Examples of these tasks include classifying galaxies and transcribing textual content from images~\cite{law2011human,lintott2013human,ponciano2014volunteers}. Human computation systems gather a crowd of people connected to the Internet and manage them to execute the tasks. The precursor of such systems is reCAPTCHA~\cite{von2008recaptcha,law2011human}. Currently, there is a broad diversity of human computation systems, such as: games with a purpose (GWAP)~\cite{von2008recaptcha}, contest sites~\cite{araujo201399designs}, online labour markets~\cite{sodre2017analysis,Difallah:2018}, and volunteer thinking systems~\cite{ponciano2014considering}. By focusing on the behaviour of volunteers in citizen science projects, we emphasise those human computation systems in which people act as volunteers and in which they are motivated to help science, gain new knowledge, and participate in real science initiatives~\cite{raddick2010galaxy,eveleigh2014designing}.

\subsection{Citizen Science Projects and Platforms}

In citizen science, each scientist or research team can build a\textit{ single-project platform} dedicated to his/her project or host his/her projects in a \textit{multi-project platform}. In the single-project approach, the platform consists of a Website built by a scientist (or a research group) to make the tasks available to the volunteers willing to contribute. In the multi-project approach, the platform hosts multiple projects from many different scientists. The platform manages the relationship between scientists and volunteers who participate in their projects. Scientists are provided with tools to build their projects on the platform. Volunteers can explore the projects available on the platform to find those that meet their interests.  

Examples of currently in-production crowd-sourced citizen science platforms are Zooniverse, and Crowdcrafting. Simpson et al. analyse the Zooniverse platform focusing on its software architecture~\cite{simpson2014zooniverse}. The study does not pay attention to the dynamics of multiples projects hosted on the same platform. Ponciano et al. propose a conceptual framework for reasoning about the general structure of multi-project platforms~\cite{ponciano2014considering}. This framework emphasises the design and management of projects, including strategies such as application composition to build projects, incentives and rewards to engage participants in the projects, tasks' dependency management, aggregation of task output to identify correct answers, task assignment algorithms, and fault tolerance. Yadav and Darlington discuss how scientists' and volunteers' motivation and participation influence the design of citizen science platforms. The authors also present guidelines for designing platforms as user-inspired socio-technical systems~\cite{yadav2016design}. 

Some studies have focused on analysing crowd-sourced citizen science projects separately. Their findings include: (1) most participants who joint the project contribute only once and do not return to perform more tasks, exhibiting only a helping behaviour~\cite{ponciano2014finding,boakes2016patterns}; (2) the interests of the volunteers in contribute to the projects include mainly helping science and learning about the project subject~\cite{raddick2010galaxy}; (3) the motivation of volunteers usually changes over the time of their contribution~\cite{rotman2012dynamic,eveleigh2014designing}; (4) many volunteers can be recruited via social networks and news sites~\cite{robson2013comparing}; (5) the success of citizen science projects can be measured by considering metrics as \textit{project appeal} and \textit{resource savings}~\cite{cox2015defining}.  

To sum up, previous studies clarified several aspects of the participation of volunteers in citizen science projects. However, little progress has been made to understand volunteers' participation beyond a single project in multi-project platforms, their experience on this type of platform and the dynamics of their cross-project engagement. This lack of understanding is a shortcoming because the pattern of engagement of volunteers in various projects may reveal the platform's value in connecting volunteers and scientists.

\section{Assessing Cross-Project Engagement on Platforms}

Our approach is intended to support the work of the manager of the platform. It helps the manager to assess the cross-project engagement of the volunteers. It also helps in understanding how different the volunteers behave when multiple projects are available and how volunteer behaviour impacts the recruitment and engagement capabilities of the platform. We model the ecosystem of the multi-project platforms, propose metrics for characterising the dynamics of cross-project engagement of volunteers on such platforms, and propose a communicability method for studying platforms' cross-project engagement features. 

\subsection{Multi-project Ecosystem: Scientists, Volunteers and Platform}

\begin{figure*}[ht]
  \centering
  \includegraphics[width=\linewidth]{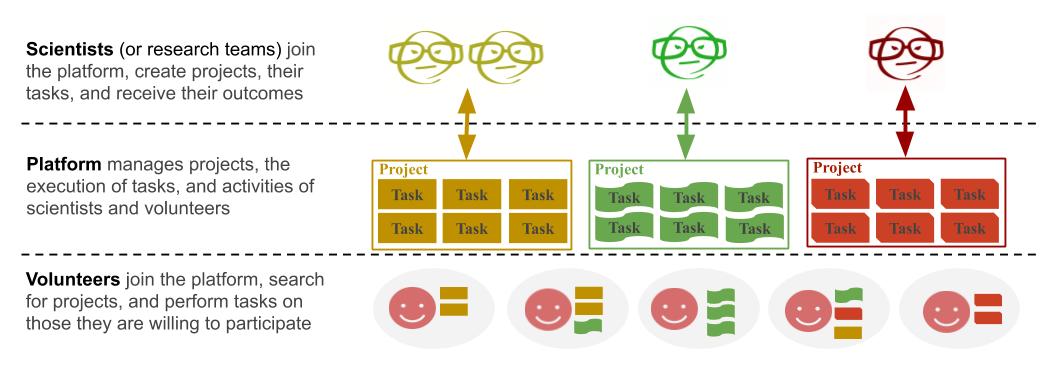}
  \caption{Projects, tasks, scientists and volunteers in the scenario of a multi-project citizen science platform. Platform is positioned to act as an intermediary between volunteers and scientist (or research teams).}
  \label{fig1}
  \Description{Overview of multi-project}
\end{figure*}

For characterising the dynamics of cross-project engagement and recruitment, we consider the three main components that make up the ecosystem built by multi-project platforms: volunteers, scientists and platform. In this ecosystem, professional scientists (or a research team) join the platform to create projects, define tasks in which volunteers can participate and receive the outcomes from them. The platform provides tools and manages the activities of scientists and volunteers. Volunteers join the platform, search for projects and perform micro tasks. Figure~\ref{fig1} summarises the roles of these components. This figure highlights that the platform is positioned to act as an intermediary between volunteers and scientists. Scientists create projects composed of micro tasks that can be performed by volunteers. Volunteers can perform micro tasks from projects of various scientists. 

\subsection{Metrics to Study Cross-Project Engagement}
Our quantitative analysis is based on the Goal, Question and Metric (GQM) approach~\cite{van1999goal}. The idea behind the GQM approach is that some explicit goals must be defined to drive the establishment of questions that will guide the definition of metrics. Based on the perspectives of the scientists, the volunteers and the platform, we derive the goals of each of them. For each goal, we define two questions that can be answered with one metric. Goals, questions and metrics are summarised in Table~\ref{tab:gqm}. The metrics were designed so that they can be used by platform managers to elicit behaviours from volunteers and to analyse those who exhibit multi-project contributing behaviour. 

\begin{table*}
  \caption{Goals, Question and Metric approach of multi-project citizen science platforms}
  \label{tab:gqm}
  \begin{tabularx}{\linewidth}{cXXX}
    \toprule
    Perspective&\multicolumn{1}{c}{Goal}&\multicolumn{1}{c}{Question}&\multicolumn{1}{c}{Metric}\\
    \midrule
    \multirow{2}{*}{Volunteer}      &  \multirow{2}{\hsize}{To discover new relevant projects in which they can contribute and sustainably engage with them.} & To what extent does the volunteer experience the diversity of projects available on the platform? & Exploration rate  \\ \\
         &   &  To what extent does the volunteer regularly engage with the projects available on the platform? & Engagement rate\\ \\
         &   &  Do multi-project volunteers stay on the platform longer than single-project volunteers? & Relative activity duration\\ \\ \hline
     \multirow{2}{*}{Scientist (or research team)} & \multirow{2}{\hsize}{To recruit new volunteers to their project and engage existing ones to have the tasks of the project executed.} & To what extent does the project recruit more volunteers outside the platform than volunteers who were already participating in the platform? & Balance in recruitment\\ \\
         &   & To what extent do volunteers recruited by the project outside the platform perform more tasks in the project than those volunteers inherited from other projects on the platform? &Balance in computing\\ \\ \hline
    \multirow{2}{*}{Platform} & \multirow{2}{\hsize}{To build a sustainable citizen science community that benefits both scientists and volunteers.} & Do few projects recruit most volunteers to the platform? & Inequality in volunteers' recruitment by the projects\\ \\
        &    & Do few projects receive the most of the contribution of the volunteers regarding the amount of performed tasks? & Inequality in the received contribution by the projects  \\
  \bottomrule
\end{tabularx}
\end{table*}

The metrics are based on data of task execution by volunteers. For each task, the data kept includes the time in which the task was performed, the unique identifier of the volunteer who performed the task and the unique identifier of the project to which the task is associated. Typically, in citizen science platforms, this data is generally maintained in logs. The use of logs eases the generation of the metrics by the platform managers. The results attained by the metrics and methods are sensitive to the context of the platform in terms of the nature of the tasks, the number of hosted projects, and the characteristics of the projects, and user interface. The proposed metrics are presented in the following for each perspective: volunteers, scientists, and platform. 

\textbf{Volunteers' perspective}: Volunteers have several reasons to participate on multi-project platforms. They may want to discover new relevant projects from which they can learn, contribute, and engage. We propose three metrics to investigate these aspects:

\begin{itemize}
    \item{\bf Exploration rate}: Let $p$ be the number of projects in which the volunteer performed at least one task and let $a$ be the number of projects available or built on the platform after the volunteer came to it. The exploration rate of the volunteer is computed as $p/a$. For example, if there are 40 projects on the platform and the volunteer executed tasks in only 10 projects, his/her exploration rate is $10/40=0.25$. Thus, exploration rate includes the temporary involvement, such as executing a task on the project out of curiosity. The closer to 1 is the value of this metric, the more the volunteer try the projects available on the platform. 
    \item{\bf Engagement rate}: Let $g$ be the number of projects in which the volunteer contributed as a regular volunteer and $a$ be the number of projects available or built on the platform after the volunteer came to it. The engagement rate is computed as $g/a$. For example, if there are 40 projects on the platform and the volunteer executed tasks in at least two different days in just 4 of these projects, his/her engagement rate is $4/40=0.1$. Thus, engagement rate reflects a lasting involvement, covering contributions that last longer than a day. The closer to 1 is the value of this metric, the more the volunteer engage regularly with the projects available on the platform. By definition, for each volunteer, the engagement rate is always less than or equal to the exploration rate.  
    \item{\textbf{Relative activity duration}}: This metric is the ratio of the number of days between the first and the last contribution of the volunteer in relation to duration of the user relationship with the platform (measured as the number of days since the volunteer joined the platform until the day when the data of the platform was collected~\cite{ponciano2014finding}). The closer to 1 is the value of this metric, the longer the volunteer has remained connected to the platform in relation to the time in which the platform was observed. This metric is used to assess if volunteers who engage in multiple projects exhibit more sustained engagement than those who engage in only one project.
\end{itemize}

\textbf{Scientists' perspective}: Scientists have several benefits when designing their projects on multi-project platforms. For example, scientists can benefit by reducing the cost of recruiting volunteers since it is likely they inherit for their projects some volunteers who already participate in other projects hosted on the platform. We define two metrics to analyse these aspects: 

\begin{itemize}
    \item{\bf Balance in recruitment}: It measures the extent to which the number of volunteers inherited from the other projects on the platform is larger than the number of volunteers recruited outside the platform. Let $n$ be the number of volunteers that the project inherited from the platform and $u$ be the number of volunteers that the project recruited outside the platform. The balance in recruitment is measured as $(n-u)/min(n,u)$. When the balance is 0, the number of recruited volunteers equals the number of inherited volunteers. When the result is positive, it says how many times the number of inherited volunteers is higher than the number of recruited volunteers. When the result is negative, it says how many times the number of recruited volunteers is higher than the number of inherited volunteers.
    \item{\bf Balance in computing}: It measures the extent which the volunteers inherited from other projects on the platform execute in average more task than those volunteers recruited by the project from outside the platform. Let $t$ be the average number of tasks performed by the inherited volunteers and $m$ be the average number of tasks performed by the recruited volunteers. The balance in computing is measured as $(t-m)/min(t,m)$. When the balance is 0, the average number of tasks performed by recruited volunteers is equal to the average number of tasks performed by inherited volunteers. When the result is positive, it says how many times the average number of tasks performed by inherited volunteers is higher than the average number of tasks performed by recruited volunteers. When the result is negative, it says how many times the average number of tasks performed by recruited volunteers is higher than the average number of tasks performed by inherited volunteers. 
\end{itemize}

\textbf{Platform's perspective}: Platforms want to build a sustainable citizen science community capable of advancing science. This objective is achieved by creating a collaborative environment from which scientists and volunteers can benefit. In doing so, it is essential that all projects help the community grow by recruiting new volunteers and help keep engaged the existing ones~\cite{ponciano2014finding,Ponciano2018social}. Also, all projects must receive some attention from volunteers to succeed. We define two metrics to analyse these aspects:
\begin{itemize}
    \item{\bf Inequality in the recruitment of volunteers} It is the Gini coefficient in the number of volunteers recruited from outside the platform by the projects. According to this metric, the maximal equality (value 0) will be one in which every project recruits the same number of volunteers; the maximal inequality (value 1), in turn, will be one in which a single project recruits 100 per cent of the total volunteers and the remaining projects recruits none.
    \item{\bf Inequality in received contribution from volunteers}  It is the Gini coefficient in the number of tasks performed by the volunteers in the projects. According to this metric, the maximal equality (value 0) will be one in which each project has the same proportional amount of performed tasks. In its turn, the maximal inequality (value 1) will be one in which a single project concentrates 100 per cent of the total executed tasks in the system and the remaining projects concentrates none.
\end{itemize}

\subsection{Inspection of Cross-Project Features}

We consider three main features implemented in platforms: to allow the participants to know the projects in which they already contributed, to allow the participants to search for new projects in which they may be willing to contribute and to allow the participants to find preferred or recommended projects by the platform. To study these features, we applied the Semiotic Inspection method~\cite{de2006semiotic}, whose theoretical basis lies in the Semiotic Engineering~\cite{de2005semiotic}. This method is based on inspection of the Web interface of the platform and do not involve observing users' interaction with systems. 

The semiotic inspection method evaluates the communicability of software. It assesses how well the designer-to-user meta-communication message gets across to the users through the interface of the systems. The application of the method consists of the analysis of meta-linguistic, static and dynamic signs; a comparison of the designer's meta-communication message codified in the signs; and a final evaluation of the communicability of the system. The method allows the identification of communication ruptures in the system and the triangulation with other methods or use cases to generate new knowledge and hypotheses. 

\section{Characterisation of Real Platforms}

In this section, we use the proposed approach to characterise real multi-project citizen science platforms. We first introduce the platforms and detail the data set collected from them. Then, we present the results on the cross-project engagement exhibited by the volunteers, and the cross-project engagement features provided by the platforms. Finally, we discuss the results, their implications, and their limitations.

\subsection{The Platforms}

Three platforms form the basis of our quantitative study: Crowdcrafting (\url{crowdcrafting.org}), Socientize (\url{socientize.eu}) and GeoTag-X (\url{geotagx.org}). We collected data about task execution events for all projects on these platforms. For each task execution event, we collected the unique identifier of the volunteer who performed the task, the unique task identifier; the identifier of the project to which the task is associated; and, the date and time that the volunteer has performed the task. Below, we detail the platforms and the data collected from each of them. 

The Crowdcrafting platform assists in performing tasks that require human cognition, knowledge or intelligence such as image classification, and transcription. From this platform, we collected data of 1,252,502 tasks execution events generated by 26,133 volunteers in the period between July 7, 2012 and July 17, 2014. The resultant dataset covers 22 projects in many fields such as the Humanities, Biology, and Arts. The Socientize platform, in turn, is a European Commission initiative to study society as infrastructure for e-science. From Socientize, we collected data of 141,808 task execution events generated by 1,667 volunteers between March 25, 2013, and May 1, 2015. The dataset includes information about ten projects from distinct subjects, including Biology and Astronomy. Finally, the GeoTag-X platform aims to help disaster relief efforts by asking volunteers to analyse photos taken in disaster-affected areas. The dataset collected from GeoTag-X covers 21,200 task execution events generated by 727 volunteers, in 16 projects, between May 2, 2014 and March 3, 2016. GeoTag-X and Socientize are currently retired. 

These platforms are run by different lead scientists, span different fields of science and host projects with different characteristics regarding what they ask volunteers to do. Despite the differences, they are developed using the same software layer called PyBossa, an open source software that eases the development of citizen science platforms~\cite{daniel_lombrana_gonzalez_2018_1402409}. The collected data refers to the period in which they were active. The data collection focused on the execution of tasks by volunteers. Thus, we collected data only from projects in which there was at least one task performed. Demonstration and test projects were not considered in the analysis. As there is no available information about the project that recruited each volunteer, we estimate that the volunteer was recruited by the project in which she/he performed the first task as soon as she/he registers on the platform. 

Data made available by the platforms have a unique identifier for the volunteers, but it is not possible to identify which individual on the platform is associated with each identifier. In this study, the volunteer data is used anonymously and confidentially, posing no risk to the volunteers and no threat to their privacy. When reporting data in the manuscript, the data are always aggregated statistically, so that no volunteer is individually known. 

In the inspection of multi-project features following the Semiotic Inspection Method~\cite{de2006semiotic} protocol, we analysed the Crowdcrafting (described previously), Zooniverse and CitSci.org platforms. Zooniverse (zooniverse.org) is a popular platform for crowd-sourced citizen science research. CitSci.org provides tools for the entire research process including gathering participant feedback and analysing collected data. These platforms are open to the creation of projects, and they host projects from several different disciplines, making them suitable for our cross-project engagement analysis. The communicability analyses were performed in the Web interface of the platforms by January 2018. The inspection scenario was the same in all platforms, considering as a user a volunteer who wants to contribute to multiple projects, maintain long-term relationships with these projects as well as discover new projects that meet their interests. The evaluators that carried out the inspection are not directly involved with the platforms and performed the test following the protocol proposed by de Souza et al.~\cite{de2006semiotic}.

The proposed quantitative and qualitative studies are complementary, but they are not directly associated with each other. The quantitative study was done using historical data of the Crowdcrafting, Socientize and GeoTag-X platforms since only these platforms made the data available publicly. We downloaded the data through their PyBossa standardised application programming interface (API). The qualitative study was carried out with the current Web interface of the platforms; thus, the study focuses on platforms that are still in-production: Crowdcrafting, Zooniverse and CitSci.org. GeoTag-X and Socientize are retired.

\subsection{Results}

As shown in Table~\ref{tab:freq}, we first analyse the distribution of volunteers who fall into the classes associated to the platform ('platform regular' and 'platform transient') and classes associated to the projects ('multi-project explorer', 'multi-project regular', and 'one project'). The platform regular class is made up of those volunteers who executed tasks on at least two days on the platform, regardless of whether it was on several projects or not. The platform transient class is made up of those volunteers who executed tasks on the platform on only one day, regardless of whether it was on several projects or not. The multi-project explorer class is made up of those volunteers who performed tasks on at least two different projects on the platform. The class of multi-project regular volunteers is made up of those volunteers who executed tasks on least two different projects on the platform. Finally, one project volunteers are those who neither explored nor regularly engaged in multiple projects. The platform regular volunteers and platform transient volunteers are mutually exclusive classes. The multi-project explorer, multi-project regulars, and 'one project' classes are also mutually exclusive.  

\begin{table*}
  \caption{Distribution of volunteers in classes considering the platform ("platform regular" and "platform transient") and projects ("multi-project explorer", "multi-project regular" and "one project") on Crowdcrafting, Socientize and GeoTag-X platforms.}
  \label{tab:freq}
  \begin{tabular*}{\textwidth}{c@{\extracolsep{\fill}}cccc}
    \toprule
    Dimension&Classes&Crowdcrafting&Socientize&GeoTag-X\\
    \midrule
    \multirow{3}{*}{Platform}      &  \multicolumn{1}{l}{Platform regular} & \multicolumn{1}{r}{$7\%$} & \multicolumn{1}{r}{$33\%$} & \multicolumn{1}{r}{$27\%$}\\
          & \multicolumn{1}{l}{Platform transient} &  \multicolumn{1}{r}{$93\%$} & \multicolumn{1}{r}{$67\%$} & \multicolumn{1}{r}{$73\%$}\\
          & \multicolumn{1}{r}{\textit{Sum}} & \multicolumn{1}{r}{$100\%$}  & \multicolumn{1}{r}{$100\%$} & \multicolumn{1}{r}{$100\%$}\\ \hline \\
          
     \multirow{4}{*}{Project}       &  \multicolumn{1}{l}{Multi-project explorer} & \multicolumn{1}{r}{$15\%$} & \multicolumn{1}{r}{$13\%$}  & \multicolumn{1}{r}{$26\%$}\\
        &  \multicolumn{1}{l}{Multi-project regular} & \multicolumn{1}{r}{$1\%$} & \multicolumn{1}{r}{$5\%$}  & \multicolumn{1}{r}{$6\%$}\\
        &  \multicolumn{1}{l}{One project} & \multicolumn{1}{r}{$84\%$} & \multicolumn{1}{r}{$82\%$}  & \multicolumn{1}{r}{$68\%$}\\
        & \multicolumn{1}{r}{\textit{Sum}} & \multicolumn{1}{r}{$100\%$}  & \multicolumn{1}{r}{$100\%$} & \multicolumn{1}{r}{$100\%$}\\
  \bottomrule
\end{tabular*}
\end{table*}

The results indicate that between 13 per cent and 26 per cent of the volunteers who join the platform are multi-project explorers. A sign that they test several projects to find those in which they want to contribute regularly. Less than a third of volunteers are platform regulars, and a small fraction (<6 per cent) are multi-project regulars. The main conclusion drawn from such distributions is that \textit{few volunteers acting on the studied platforms explore multiple projects and the minority exhibit permanent engagement either in multiple projects or a particular project}. 

\begin{figure*}
	\centering
	\subfigure[Crowdcrafting]{ 
		\includegraphics[scale=0.5]{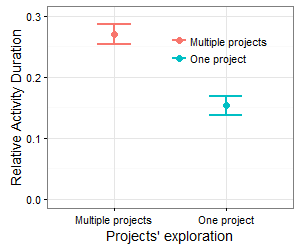}
	}
	\subfigure[GeoTag-X]{ 
		\includegraphics[scale=0.5]{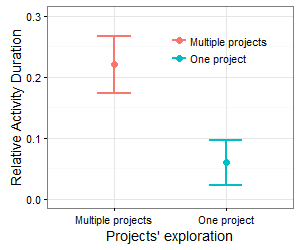}
	}
	\subfigure[Socientize]{ 
		\includegraphics[scale=0.5]{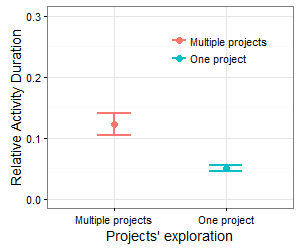}
	}
	\caption{Relative activity duration of regular volunteers who contribute to a single project and of regular volunteers who contribute to multiple projects on the Crowdcrafting, Socientize, and GeoTag-X platforms. Confidence intervals for a statistical confidence level of 95 per cent} 
	\label{fig2}
\end{figure*}

Focusing on volunteers who contributed on the platform on at least two different days, we investigate whether multi-project regular volunteers exhibit higher engagement scores than one project regular volunteers (Figure~\ref{fig2}). The results show that \textit{volunteers who regularly engage in various projects stay longer on the platform than those who engage in only one project}. Volunteers who engage in multiple projects tend to exhibit higher relative activity duration. This is evidence that volunteers who find new projects that met their interests end up staying longer on the platform.  

Turning our attention to the platform' perspective, we analyse the inequalities among projects regarding the number of recruited volunteers and the amount of contribution received from the volunteers (Table~\ref{tab:ineq}). The result indicates\textit{ a concentration of attention in a small subset of projects: a few projects attract most of the attention from the volunteers and are responsible for most of the activity of the volunteers on the platforms}. In all the studied platform, few projects recruit most of the volunteers and receive most of the contributions from them. It is the case of for example of the ``Dark Skies ISS'' project on the Crowdcrafting platform, the ``Cell Spotting'' project on the Socientize platform and the ``Yemeni Cultural Heritage at Risk'' project on the GeoTag-X platform. From Table 3, it can be noted that the GeoTag-X platform presents the lowest values of inequality among the projects. One of the contributing factors to this is the fact that this platform is more homogeneous, i.e. the projects are more like each other (they are related to image georeferencing). Therefore, increasing the chance that volunteers of the platform will have an interest in many of them. The other platforms host more different projects among themselves, and the platform volunteer may be interested in only a few of them. 

\begin{table}
  \caption{Inequalities in the number of volunteers recruited by the projects and in the number of tasks performed (Gini coefficient) on the Crowdcrafting, GeoTag-X, and Socientize platforms. The closer to 1 the higher is the inequity.}
  \label{tab:ineq}
  \begin{tabularx}{\linewidth}{cXX}
    \toprule
    Platform&Inequality in volunteers recruitment&Inequality in received contribution\\
    \midrule
    Crowdcrafting      &  \multicolumn{1}{r}{$0.93$} & \multicolumn{1}{r}{$0.95$}\\
    GeoTag-X       &  \multicolumn{1}{r}{$0.47$} & \multicolumn{1}{r}{$0.64$}\\
    Socientize    &  \multicolumn{1}{r}{$0.61$}    & \multicolumn{1}{r}{$0.80$}\\
  \bottomrule
\end{tabularx}
\end{table}

\begin{figure*}
	\centering
	\subfigure[Crowdcrafting]{ 
		\includegraphics[scale=0.75]{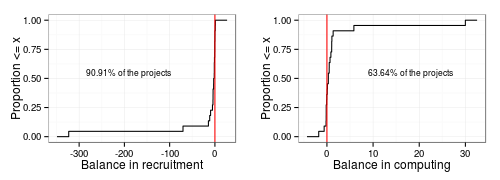}
	}
	\subfigure[GeoTag-X]{ 
		\includegraphics[scale=0.75]{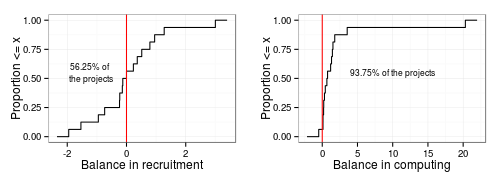}
	}
	\subfigure[Socientize]{ 
		\includegraphics[scale=0.75]{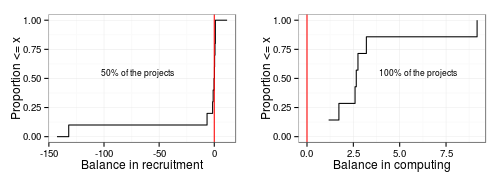}
	}
	\caption{Empirical Cumulative Distribution Function of the balance in the recruitment of new volunteers (right) and in the balance in computing (left) on Crowdcrafting, GeoTag-X and Socientize platforms. The vertical lines show the equality, which corresponds to the case in which the balance is 0. A positive balance indicates that the project has benefited from the platform more than contributed to it.} 
	\label{fig3}
\end{figure*}

Finally, we analyse the perspective of the scientist or research team. Considering the balance in recruitment, the results of the three platforms show that the proportion of projects that exhibit a negative balance in computing is equal to or larger than 50 per cent (Figure~\ref{fig3}).  

A negative balance in recruitment indicates that a project recruited more volunteers than inherited from the platform. Considering the balance in computing, the results show that most of the projects exhibit a positive balance. A positive balance in computing indicates that the average number of tasks performed by volunteers inherited from the platform is larger than the average number of tasks performed by volunteers recruited from outside the platform (Figure~\ref{fig3}). The conclusion we can draw from these results is that \textit{new projects inherit fewer volunteers from the platform than they recruit, but such inherited volunteers perform more tasks than the recruited ones}. 

To gain insight into what is provided to the volunteers by the platforms so that they can engage with multiple projects, we analyse the interface of three multi-project platforms: Zooniverse, Crowcrafting and CitSci.org. As discussed earlier, we used the semiotic inspection. We identify three classes of signs in platforms' interfaces. 

\begin{itemize}
    \item {\bf Signs of project search}:  The platform displays these signs on a specific page dedicated to project search. These pages are usually accessed via a drop-down menu item labelled as ``Projects'' (Zooniverse and CitSci.org) or ``Discover'' (Crowcrafting). The page contains fields for the insertion of terms to be searched in project names. It guides the user in defining search terms and filters the results by fields such as most recently launched and most active projects. Zooniverse also allows volunteers to filter by the state of the project: Active, Paused, and Finished. The user must know filter's semantics to use them. These signs are visualised even when the user is not logged into the platform.
    \item {\bf Signs of projects preferred or recommended by the platform}: They are signs that present projects that the user should know, try or whose prominence should be known. It is usually referred to as "Featured Project(s)". The number of featured projects varies from one to several. They are identified by an image and their names. On all of the studied platforms, the criteria used to make a ``featured'' project are unclear. Also, the platforms do not recommend to the volunteer other projects that are related to those in which they already have contributed or that are related to their task execution behaviour.
    \item {\bf Signs of participant's projects}: They are usually shown in the user profile page or in the first page shown as soon as the user logs in to the platform. ``My projects'', ``Your contributions'' or ``My recent projects'' are signs used to inform users about the projects in which they have participated. The projects are usually identified by an image that represents them and their names. These signs are visualised only with the user logged into the platform.
\end{itemize}

\subsection{Discussions and Recommendations}

In this work, we propose and evaluate a conceptual and methodological framework to study volunteer recruitment and engagement in the context of crowd-sourced multi-project citizen science platforms. By using the framework, we characterise volunteers' cross-project engagement patterns and inspect the communicability of platforms' cross-project features in real platforms. Now we turn to detail recommendations and insightful reflections derived from our findings. 

{\bf Inequalities in volunteers' cross-project engagement.} Our results suggest that most volunteers do not try multiple projects on the platforms. The volunteers who try multiple projects tend to stay active on the platform for longer, and a small fraction of them ends up regularly participating in more than one project. These results indicate that the inequality in engagement patterns observed when analysing a single project are also observed when analysing an entire platform composed of multiple projects. The widely discussed 1 per cent rule of thumb pertaining to engagement in online systems~\cite{ponciano2014finding}. 

{\bf Inequalities in projects ability to recruit, engage, and attract the attention of volunteers.} A small subset of projects keeps the platform active in terms of recruiting volunteers and receiving contribution from them. Also, a new project tends to recruit more volunteers to the platforms than it inherits from them, but such inherited volunteers are more engaged than the recruited ones. One reason for this result is that recruitment campaigns put into practice by the owners of new projects usually consists of a broadcasting campaign in social networks~\cite{robson2013comparing}. This kind of campaigns have the potential to attract many volunteers, but most of them are just curious and do not exhibit long-term engagement~\cite{ponciano2014finding}. Moreover, volunteers inherited from other projects on the platform tend to be more engaged because they have been shown to be intrinsically motivated exhibiting volunteering behaviour instead of helping behaviour~\cite{clary1998understanding,wilson2000volunteering}.

\textbf{Features provided to volunteers interact across projects on the platform.} The semiotic analysis of signs used by platforms to indicate their cross-project engagement features shows that platforms focus mainly on allowing volunteers (1) to keep a record of the projects in which they have already contributed, (2) to find new projects in which they may be willing to contribute, and (3) to find projects preferred or recommended by the platform. Such features are usually associated with a global composition of the platform. For example, platforms allow volunteers to search for projects, but, to use this feature, the volunteer must know the name of the project, making it challenging to explore unknown topics. Also, the recommendation of projects as ``featured projects'' is associated with the platform as a whole and not a personalised characteristic of each volunteer.

\textbf{Implications for design.} We extract three major guidelines from our results. 
\begin{enumerate}
    \item Platforms should encourage a cross-project engagement when the volunteer is exhibiting an explorer behaviour, when he/she has not permanently engaged with a project or when such a project has been completed. In such situations, engaging in multiple projects can help prevent the volunteer from leaving the platform.
    \item Projects' recommendations should be personalised to each volunteer according to their participation preferences and behaviour, and explanations should be provided about why that is a suitable recommendation, in a way to be transparent and educative. For example, platforms should recommend to the volunteers other projects that are related to those in which they have already have contributed. As the number of projects on the platform increases and projects tend to be very different from each other, this kind of functionality becomes essential in helping the volunteer find projects that are relevant to her/him.
    \item Platforms should provide feedback and recognition to the volunteers for their multi-project participation. Since the number of volunteers participating in multiple projects on the platforms is not negligible, this participation pattern should be analysed and recognised when it is beneficial. 
\end{enumerate}

\textbf{Implications for research.} One important new question raised in our study is what the role of the platform and the projects in the relationship with the volunteers should be. For example, should engagement strategies be put into practice by the project or by the platform? In email and social network campaigns, the project tends to ask the volunteer to return to perform more tasks, while the platform tend to encourage volunteers to return to contribute to a new, different project. It is fundamental to enhance the quality of participation and to encourage a sustainable engagement per project, but it is also beneficial the fact volunteers try multiple different projects. In doing so, they can discover new relevant projects in which they may be willing to contribute and learning from. It remains a challenge to identify how platforms and projects can have their roles defined to promote a win-win-win situation in which platform, scientists, and volunteers can benefit even more from citizen science.  

\bigskip

\textbf{Limitations.} As discussed earlier, the scope of the study are crowd-sourced citizen science platforms implemented as Web-based systems, which host multi-projects, and in which volunteers are free to join and contribute by performing tasks based on cognitive processing. The proposed communicability inspection approach builds on semiotic inspection, which has been used in Web-based systems in general. The GQM approach and the proposed metrics rely on task execution data that is typically stored by the citizen science platforms; so, by construction, they are not based on data which are available only on specific platforms. However, the results of the use of the methods and metrics on a specific platform cannot be generalised to other platforms. Metrics based on the number of performed tasks can become inaccurate if the tasks differ greatly in complexity, which is not the typical case in the micro-task context. Our empirical analyses are case studies and cannot generalizable to systems that fit into a domain other than those studied. Our methods and metrics are proposed to help designers and managers when analysing a platform, but not to perform a cross-platform comparison, in which several other aspects should be considered such as project numbers, type of tasks and way of recruiting volunteers.  

\section{Conclusion}

We sought to characterise how volunteers behave across projects on multi-project citizen science platforms and the benefits they take from them. In doing so, we provided three major contributions: 1) we integrated a set of concepts and theories in a framework for analysing multi-project platforms from the perspective of human engagement, volunteering and design of platforms; 2) using a GQM model, we derived metrics to assess underlying properties of volunteers, scientists and platforms; 3) we characterise cross-project engagement on three platforms and inspect the communicability of cross-project features implemented on three platforms. 

Our results show several characteristics from volunteers, scientists, and their interaction on the platforms. We discuss such characteristics highlighting that volunteers engage little in multiple projects, there is significant inequality in the attention projects receive from volunteers, recruiting volunteers from other projects is a positive thing for both projects and volunteers. Our empirical analyses let us identify and highlight opportunities for improving the designing and management of multi-projects platforms. Cross-platform analysis of volunteer engagement and recruitment is an important topic for future work. Task types, duration, technologies, media exposure and training are examples of important aspects to consider. We hope the methods and findings that we report in this study motivate new discussions and studies in this direction. 

\begin{acks}

We are indebted to the platforms that made available the datasets used in this study. We are also grateful to Francisco Brasileiro, Nazareno Andrade, and the anonymous reviewers for their valuable comments and feedback, which contributed to improve our work.
\end{acks}

%
\bibliographystyle{ACM-Reference-Format}
\bibliography{sample-base}

\end{document}